\documentclass[%
reprint,
superscriptaddress,
amsmath,
amssymb,
aps,
prl,
floatfix,
]{revtex4-2}

\usepackage{graphicx}
\usepackage{dcolumn}
\usepackage{bm}
\usepackage[hypertexnames=false]{hyperref}
\hypersetup{
	colorlinks = true,
	allcolors = blue,
}

\usepackage{multirow}

\begin{document}

\title{Shell-shaped Bose-Einstein condensates realized with dual-species mixtures}

\author{A. Wolf}
\email[Corresponding author. Email: ]{a.wolf@dlr.de}
\affiliation{Institute of Quantum Technologies, German Aerospace Center (DLR), 89077 Ulm, Germany}
 
\author{P. Boegel}
\affiliation{Institut f\"ur Quantenphysik and Center for
	Integrated Quantum Science and Technology ($\it IQ^{ST}$), Universit\"at Ulm, 89081 Ulm, Germany}

\author{M. Meister}
\affiliation{Institute of Quantum Technologies, German Aerospace Center (DLR), 89077 Ulm, Germany}

\author{A. Balaž}
\affiliation{Institute of Physics Belgrade, University of Belgrade, 11080 Belgrade, Serbia}

\author{N. Gaaloul}
\affiliation{Institut f\"ur Quantenoptik, Leibniz Universit\"at Hannover, 30167 Hannover, Germany}

\author{M.A. Efremov}
\affiliation{Institute of Quantum Technologies, German Aerospace Center (DLR), 89077 Ulm, Germany}
\affiliation{Institut f\"ur Quantenphysik and Center for
Integrated Quantum Science and Technology ($\it IQ^{ST}$), Universit\"at Ulm, 89081 Ulm, Germany}

\date{\today}

\begin{abstract}
Ultracold quantum gases confined in three-dimensional bubble traps are promising tools for exploring many-body effects on curved manifolds. As an alternative to the conventional technique of radio-frequency dressing, we propose to create such shell-shaped Bose-Einstein condensates in microgravity based on dual-species atomic mixtures and we analyze their properties as well as the feasibility to realize symmetrically filled shells. Beyond similarities with the radio-frequency dressing method as in the collective-excitation spectrum, our approach has several natural advantages like the robustness of the created quantum bubbles and the possibility to magnify shell effects through an interaction-driven expansion.
\end{abstract}
\maketitle
\noindent\textit{Introduction}.--
Performing many-body physics on a shell topology opens a new avenue in studying nontrivial quantum phenomena, such as Bose-Einstein condensation~\cite{Dalfovo.1999,Tononi.2019,Moller.2020,Tononi.2020,Rhyno.2021}, atom lasers~\cite{Robins.2013,Meister.2019b}, superfluidity~\cite{Pitaevskij.2016}, vortices~\cite{Fetter.2009,Turner.2010,Padavic.2020,Bereta.2021} and the Berezinskii-Kosterlitz-Thouless transition~\cite{Kosterlitz.2016,Tononi.2021}. Indeed, quantum gases confined to the surface of a sphere show lower condensation temperatures than their filled counterparts and the thin-shell transition between a three-dimensional and quasi-two-dimensional geometry drastically changes the collective excitations~\cite{Padavic.2017,Sun.2018} and the dynamics of vortex-antivortex pairs. Nowadays, radio-frequency (rf) dressing is the primary technique to realize experimentally shell-shaped Bose-Einstein condensates (BECs) in microgravity. This technique, proposed by Zobay and Garraway~\cite{Zobay.2001,Zobay.2004,Lesanovsky.2006,Garraway.2016,Perrin.2017}, relies on the adiabatic deformation of a typically anisotropic static magnetic trap by applying a radio-frequency field. Although rf-dressing has been successful in creating a number of novel topologies and applications for BECs~\cite{Colombe.2004,Merloti.2013,Guo.2020,Guo.2021,Schumm.2005,Lesanovsky.2006b,Hofferberth.2006,Hofferberth.2007,Lesanovsky.2007,Heathcote.2008,Sherlock.2011,Harte.2018,Barker.2020}, a three-dimensional hollow sphere of atoms is beyond the capabilities of Earth-based laboratories due to gravity pulling the atoms towards the bottom of the trap~\cite{Colombe.2004}. However, the recent progress in the development of microgravity BEC machines including NASA's Cold Atom Lab (CAL)~\cite{Elliott.2018,Aveline.2020} and the Bose-Einstein Condensate and Cold Atom Laboratory (BECCAL)~\cite{Frye.2021} promises to make shell-shaped BECs experimentally feasible~\cite{Lundblad.2019,Carollo.2021}. Nevertheless, it is still a challenging task, because any inhomogeneity of the rf-field and non-perfect spatial alignment with respect to the static field potentially open the shell up, similar to gravity on Earth.

As an alternative to rf-dressing, in this Letter we propose to realize shell-shaped BECs employing a mixture of two atomic species. If the repulsive inter-species interaction outweighs the repulsive intra-species one, the mixture separates into two domains, each containing solely one type of atoms~\cite{Pethick.2008,Colson.1978}. Confining such a mixture in a three-dimensional harmonic trap leads to a regime where one species forms a shell around the other one~\cite{Ho.1996,Pu.1998b,Riboli.2002}. The scheme can be realized with an optical dipole trap~\cite{Grimm.2000} to confine the atoms and a homogeneous magnetic field to tune the atom-atom interaction via Feshbach resonances~\cite{Timmermans.1999,Chin.2010}. This realization has several advantages: (i) the atoms condense into the shell-shaped ground state instead of being adiabatically deformed into it, (ii) a homogeneous Feshbach field is much easier realized than combining multiple magnetic fields for rf-dressing, (iii) spherical symmetry of the atom cloud can be straightforwardly achieved by combining three optical trapping beams, and (iv) expanding shells are created by simply turning off the common trap because an inwards expansion of the outer species is prevented by the core one.

Here we identify the parameters required to realize shell-shaped ground states and investigate the transition signature from a filled sphere into a shell-shaped BEC by studying the collective excitation spectrum of the mixture. Moreover, we find two distinct scenarios of the free expansion depending on the inter-species interaction. Finally, we discuss the feasibility and robustness of the created quantum bubbles realized with the proposed method and contrast them with the rf-dressing ones. 

\noindent\textit{Shell-shaped ground states}.--
For a BEC well below the critical temperature, its properties can be described by a mean-field approach, leading to the Gross-Pitaevskii equation (GPE)~\cite{Ueda.2010,Pitaevskij.2016} for the condensate wave function $\psi$. In the case of BECs containing multiple components $\alpha=1,2,\ldots$ (e.g., different atomic species), the GPE for component $\alpha$ reads~\cite{Pethick.2008}
\begin{equation}
i\hbar \frac{\partial \psi_\alpha(\mathbf{x},t)}{\partial t} = \left[h_\alpha(\mathbf{x}) + \sum_\beta g_{\alpha\beta}|\psi_\beta(\mathbf{x},t)|^2\right] \psi_\alpha(\mathbf{x},t).
\label{eq:GPE}
\end{equation}
Here $h_\alpha(\mathbf{x}) = - \hbar^2 \nabla_\mathbf{x}^2/(2m_\alpha) + V_\alpha(\mathbf{x})$ is the single particle Hamiltonian of an atom with mass $m_\alpha$ and $V_\alpha(\mathbf{x})$ is the component-dependent external potential. The sum over all components in Eq.~\eqref{eq:GPE} contains the self-interaction of the component ($\beta = \alpha$) and the interaction between two components ($\beta \neq \alpha$). The interaction parameters $g_{\alpha\beta} = 2 \pi \hbar^2 a_{\alpha\beta} (m_\alpha + m_\beta)/(m_\alpha m_\beta)$ are determined by the $s$-wave scattering lengths $a_{\alpha\beta} = a_{\beta\alpha}$. The condensate wave function $\psi_\alpha$ is normalized to the number of particles $N_\alpha = \int \mathrm{d}^3x |\psi_\alpha(\mathbf{x},t)|^2$.

To create a shell-shaped BEC, we propose to use a two-component BEC mixture in a spherically symmetric harmonic confinement $V_\alpha(\mathbf{x}) = m_\alpha \omega_{0,\alpha}^2 \mathbf{x}^2/2$ with the trap frequency $\omega_{0,\alpha}$, created by an optical dipole trap~\cite{Grimm.2000}. Using magnetic Feshbach resonances~\cite{Timmermans.1999,Chin.2010}, we require that $g_\text{12} \geq \sqrt{g_\text{11} g_\text{22}}$ with $g_{\alpha\alpha} > 0$. In this phase-separation regime the inter-component repulsion outweighs the intra-component repulsion. The system therefore favors a separation of the components and reduces their overlap~\cite{Colson.1978}. Combined with the harmonic confinements, the ground state of the coupled GPEs~\eqref{eq:GPE} is given by one component forming a shell with the other one as its core~\cite{Ho.1996,Pu.1998b,Riboli.2002}.
\begin{figure}[t]
	\centering
	\includegraphics[width=245.99109pt,height=117.59764pt]{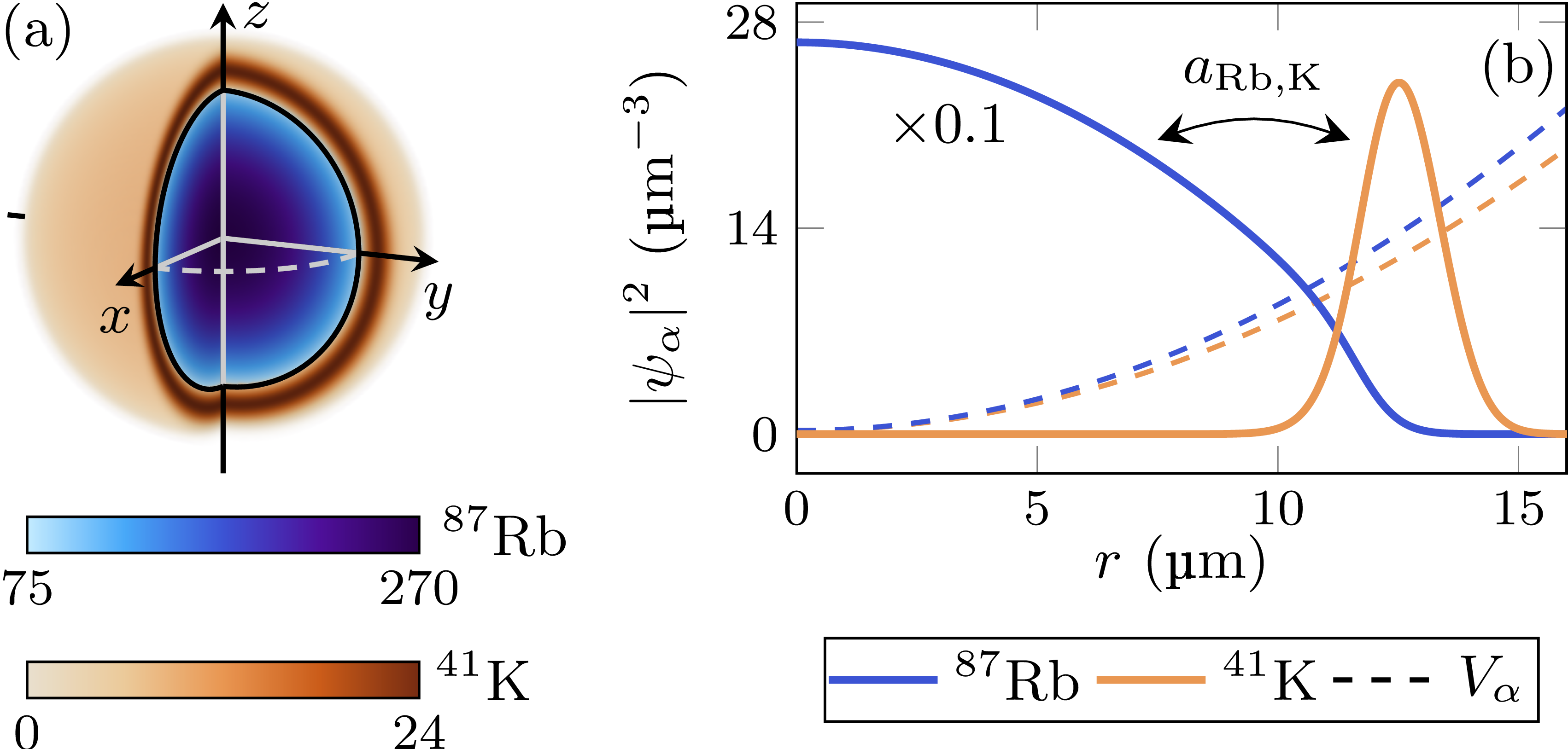}
	\caption{
	    Shell-shaped ground-state density distribution $|\psi_\alpha|^2$ of a spherically symmetric $^{87}$Rb-$^{41}$K BEC mixture for the parameters presented in Tab.~\ref{tab:reference}. (a) Cut-open three-dimensional density plot (colorbar units in $\text{\textmu m}^{-3}$). (b) Density profiles (solid lines) along the radial direction and corresponding trapping potentials $V_\alpha$ (dashed lines). The interplay between the harmonic confinement and the inter-species repulsion, due to a positive $s$-wave scattering length $a_\text{Rb,K}$, leads to $^{41}$K (orange) forming a shell around $^{87}$Rb (blue).
	}
	\label{fig:groundStates}
\end{figure}
\begin{table}[b]
	\caption{\label{tab:reference}%
	    Parameters of our reference case, based on an optically trapped $^{87}$Rb-$^{41}$K BEC mixture with laser wavelength of $1064$~nm and exploiting a Feshbach resonance at $78.9$~G to tune the inter-species interaction. Here $a_0 = 5.29\times10^{-11}$~m is the Bohr radius.
	}
	\begin{ruledtabular}
		\begin{tabular}{lclcc}
			\multicolumn{1}{c}{Species} & $N_\alpha$ & \multicolumn{1}{c}{$\omega_{0,\alpha}$~(Hz)} & $a_{\alpha\alpha}$~($a_0$) & $a_{\text{Rb,K}}$~($a_0$)\\\colrule
			\rule{0pt}{3ex }$^{41}$K & $10^5$ & $2\pi \times 70.0$ & 60 &\multirow{2}{*}{85}\\
			$^{87}$Rb & $10^6$ & $2\pi \times 51.3$ & 100 &
		\end{tabular}
	\end{ruledtabular}
\end{table}

Here we consider the parameters listed in Tab.~\ref{tab:reference} as the reference case of our analysis. Choosing $^{87}$Rb and $^{41}$K is inspired by the upcoming BECCAL apparatus~\cite{Frye.2021}, which will provide optically trapped BEC mixtures of these species in microgravity onboard the International Space Station. For $^{87}$Rb-$^{41}$K mixtures, there is a magnetic Feshbach resonance at $78.9$~G~\cite{Thalhammer.2008}, around which the inter-component scattering length $a_\text{Rb,K}$ can be tuned to a great extent whereas the intra-component scattering lengths $a_\text{Rb,Rb}$ and $a_\text{K,K}$ are kept constant at their background values~\cite{Chin.2010,Lysebo.2010}. Thus, $a_\text{Rb,K}$ is a single and well-controlled parameter to obtain a large variety of ground states. To create shell-shaped ground states, there are lower $a_\text{Rb,K}^\text{min}$ and upper $a_\text{Rb,K}^\text{max}$ limits for $a_\text{Rb,K}$. Indeed, the phase-separation regime only occurs for $g_\text{12} \geq \sqrt{g_\text{11} g_\text{22}}$, giving rise to $a_\text{Rb,K}^\text{min}\approx 72~a_0$~\footnote{This lower bound is only approximate. For $a_\text{Rb,K} < a_\text{Rb,K}^\text{min}$ the ground state is shell-shaped but the components increasingly overlap. Nevertheless, the inequality for the phase-separation regime is a good tool to estimate a lower threshold as it depends only on the masses and scattering lengths.}. However, for $g_{12} \gg \sqrt{g_{11}g_{12}}$, any contact surfaces become energetically unfavorable, resulting in a different type of ground state with side-by-side components~\cite{Trippenbach.2000}. Using the parameters of our reference case and performing three-dimensional simulations~\cite{Vudragovic.2012,Sataric.2016} of the coupled GPEs~\eqref{eq:GPE} for increasing $a_\text{Rb,K}$, we have located this transition to asymmetric ground states at $a_\text{Rb,K}^\text{max}\approx 118~a_0$. Thus, for $72~a_0 \lesssim a_\text{Rb,K} \lesssim 118~a_0$, a shell-shaped mixture is realized as displayed in Fig.~\ref{fig:groundStates}, where potassium (orange) forms a shell around rubidium (blue).

\noindent\textit{Hollowing transition and collective excitation spectrum}.--
Decreasing the scattering length $a_\text{Rb,K}$ below $72~a_0$ leads to an increasing overlap between the two components. In other words, the repulsion between $^{87}$Rb and $^{41}$K atoms becomes insufficient for  $^{87}$Rb to push $^{41}$K out of the center of the system, resulting in a non-vanishing particle density of $^{41}$K at the center. We call this transition of $^{41}$K between a filled and a hollow ground state the \emph{hollowing transition}, in analogy to rf-dressed BECs~\cite{Padavic.2017,Sun.2018}. It is a simple realization of a topological transition, resulting from the appearance of an inner surface in the $^{41}$K density.

We expect signatures of the hollowing transition when monitoring the dynamics of the mixture and therefore study its response to perturbations of the ground state. If perturbed in a sufficiently small manner, a BEC reacts linearly and oscillates in the trap with different low-lying collective excitation modes~\cite{Ueda.2010,Pitaevskij.2016}. To obtain the corresponding spectrum, we solve the Bogoliubov-de Gennes equations~\cite{Dalfovo.1999,Pu.1998,Gordon.1998,Ohberg.1999} and perform complementary simulations of the GPEs~\eqref{eq:GPE} \footnote{See Supplemental Material at [URL will be inserted by publisher] for more details on the Bogoliubov-de Gennes equations, numerical simulations, and a comparison of the mixture and rf-dressing.}.

\begin{figure}[ht]
	\centering
	\includegraphics[width=245.99658pt,height=162.42366pt]{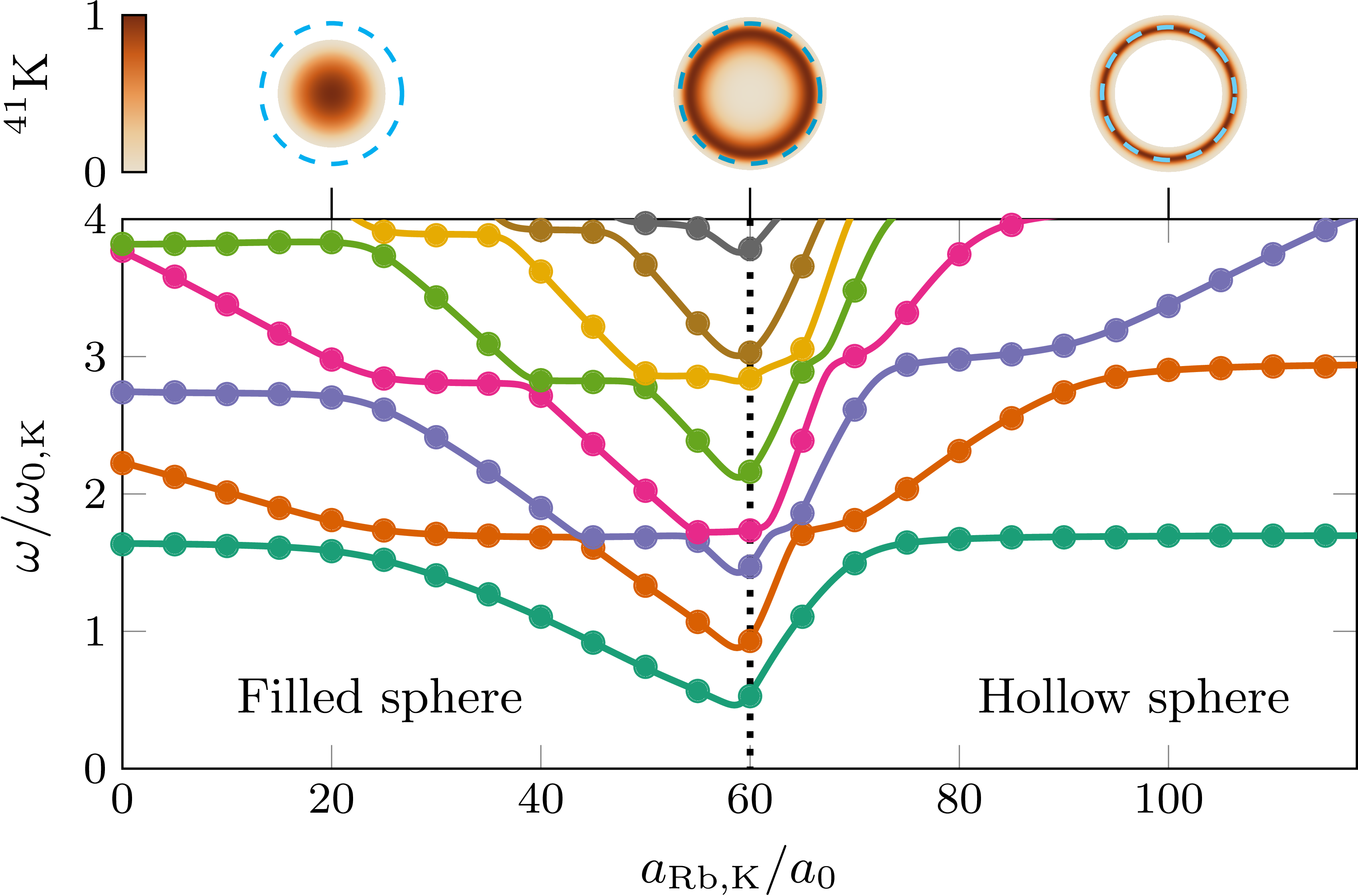}
	\caption{
	    Mode frequency $\omega$ of the lowest-lying spherically symmetric ($l=0$) collective excitations of the whole system as a function of the inter-species scattering length $a_\text{Rb,K}$. The solid lines and dots are determined by the solutions of the Bogoliubov-de Gennes equations and numerical simulations of the GPEs~\eqref{eq:GPE}, respectively~\cite{Note2}. The common minimum of the frequencies is a clear sign of the hollowing transition marked by the dotted vertical line, where $|\psi_\text{K}(0)|^2/\mathrm{max}|\psi_\text{K}(\mathbf{x})|^2$ drops below $10^{-2}$. Top: Two-dimensional cuts of the ground-state density $|\psi_\text{K}|^2$ corresponding to the marked values of $a_\text{Rb,K}$ (colormap scaled to respective peak density) illustrating the hollowing transition. The surface of $^{87}$Rb, where $|\psi_\text{Rb}|^2$ drops below $10^{-2}$ of its peak density, is indicated by the dashed blue lines. 
	}
	\label{fig:breathing}
\end{figure}

A key signature of a BEC changing its ground state topology from a filled to a hollow sphere has been identified in the spherically symmetric ($l=0$) collective excitations of an rf-dressed BEC~\cite{Padavic.2017,Sun.2018}. The corresponding frequencies show a minimum at a certain detuning related to the point of the hollowing transition. The excitation spectrum of a BEC mixture displays a similar feature. Using our reference case, for which the hollowing transition is driven by a change of the inter-component scattering length $a_\text{Rb,K}$, we see a significant fraction of mode frequencies having a minimum at $a_\text{Rb,K} \approx 60~a_0$, as shown in Fig.~\ref{fig:breathing}.
This value coincides with the critical value at which the $^{41}$K ground state vanishes in the center of the system (vertical dotted line). The excitation spectrum of the mixture helps therefore to identify a key signature of the hollowing transition and a subsequent regime of shell-shaped ground states.

A striking difference between an rf-dressed BEC and the mixture is the presence of avoided crossings in the spectrum shown in Fig.~\ref{fig:breathing}, which can be traced back to the collective excitations of the inner core-component. In our reference case, the disparity of the particle numbers leads to the ground state of $^{87}$Rb barely changing when increasing the interaction between the components. Consequently, we see excitation frequencies which are almost independent of $a_\text{Rb,K}$ as well as those which tend towards or away from the minimum. At each avoided crossing, the modes exchange the dominant component in the corresponding density oscillations~\cite{Ohberg.1999} and thus the behavior of their frequency as a function of $a_\text{Rb,K}$.

\begin{figure}[ht]
	\centering
	\includegraphics[width=245.94775pt,height=166.02522pt]{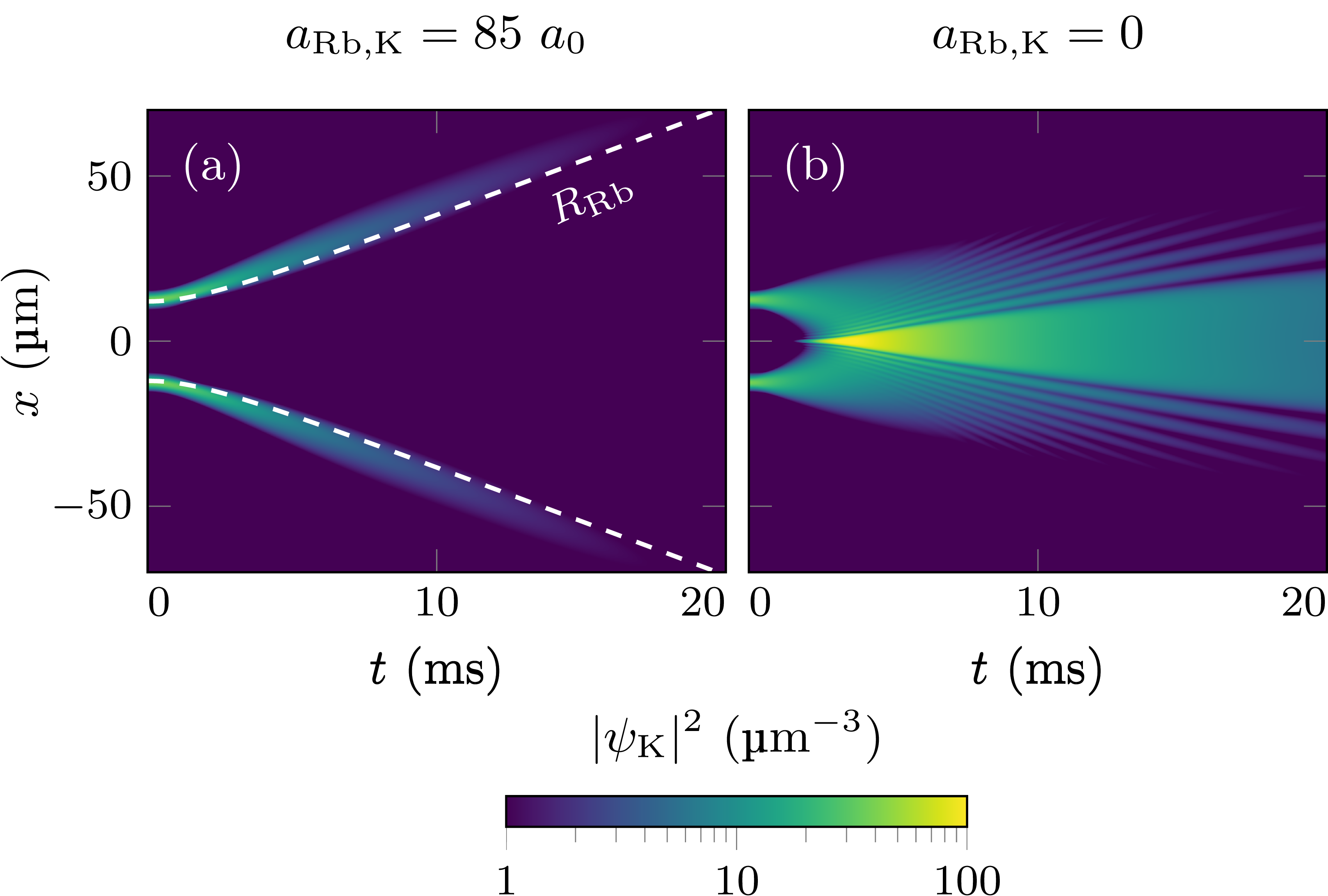}
	\caption{
	    Time evolution of the spherically symmetric density distribution $|\psi_\text{K}|^2$ along the $x$-direction for different free expansion scenarios using the initial shell-shaped state of Fig.~\ref{fig:groundStates}. (a) Solely switching off the external confinement leads to an expanding shell with its size being proportional to the edge of the expanding inner rubidium core $R_\text{Rb}$ defined by $|\psi_\text{Rb}|^2$ dropping below $10^{-2}$ of its peak value. (b) By additionally switching off the interaction between the two species at $t=0$, the shell can expand inwards until it reaches the center and shows a self-interference pattern.
	}
	\label{fig:expansion}
\end{figure}

\noindent\textit{Free expansion}.--
A special feature of mixture-realized shells is exhibited in their free expansion after the confinement is switched off. Two very distinct expansion scenarios are possible due to the freedom of controlling the inter-component interaction. Taking a shell-shaped ground state and solely switching off the harmonic confinement for both components leads to an expanding shell as displayed in Fig.~\ref{fig:expansion}(a).
Here the persisting repulsive inter-component interaction leads to the outwards expansion of the inner component ($^{87}$Rb), preventing an inwards expansion of the outer component ($^{41}$K) and therefore the shell structure is conserved. A second scenario occurs, if the inter-component scattering length can be tuned to zero at $t=0$. In this case, the two components evolve independently and the outer component can expand inwards until it reaches the center and a self-interference pattern emerges, as shown in Fig.~\ref{fig:expansion}(b).

Conserving the shell structure during the expansion is an important feature of the mixture compared to rf-dressed shells, where the typical expansion scenario is similar to the one presented in Fig.~\ref{fig:expansion}(b)~\cite{Lannert.2007,Tononi.2020}. In case of Fig.~\ref{fig:expansion}(a), after $t=20~\text{ms}$, the shell has a radius $\langle r \rangle_\text{K} \approx 80~\text{\textmu m}$ and a width $\sqrt{\langle r^2\rangle_\text{K}-\langle r\rangle_\text{K} ^2} \approx 8~\text{\textmu m}$. The natural occurrence of expanding shells presents a clear advantage of our proposed scheme, as they offer the possibility of magnifying dynamical effects like vortex formation and collective excitations on the shell surface.

\noindent\textit{Feasibility of generating shell-shaped BECs}.--
Finally, we discuss the feasibility of achieving symmetrically filled shells in the two schemes based on dual-species mixtures and rf-dressing. Shell-shaped BECs with spherically or cylindrically symmetric ground state densities are hard to generate experimentally~\cite{Zobay.2001,Colombe.2004,Hofferberth.2006, Perrin.2017,Lundblad.2019,Carollo.2021}. System-dependent effects, e.g., due to gravity or the magnetic field setup, can deform the ground state. In general, they tilt the net shell-creating potential towards one direction which is consequently favored by the atoms. Although deformations due to small tilts can be compensated by inter-atomic repulsion, if the difference in the potential minima is of the order of the chemical potential, the shell opens up completely.

Indeed, gravity is a major obstacle for creating shells with mixtures and is modeled by including an additional potential $V_{g,\alpha}(z) = m_\alpha g z$ into the single particle Hamiltonian $h_\alpha$ of each component. Here $g$ denotes the gravitational acceleration. As a result, each total single particle potential retains its harmonic form, but has its minimum shifted by $g/\omega_{0,\alpha}^2$ along the $z$-axis. A differential shift $g (1/ \omega_{0,\alpha}^2 - 1/ \omega_{0,\beta}^2)$ displaces the two components from each other, thereby compromising the symmetry of the shell. Two solutions offer themselves: (i) working in a microgravity environment to reduce $g$ or (ii) realizing equal trap frequencies for both components to achieve equal potential shifts. In our reference case, we exploit the first option and study the degree of microgravity required for closed shells. In Earth-based laboratories and following option (ii), one could realize shell-shaped BECs by carefully choosing the laser wavelength of the optical trap~\cite{Safronova.2006,Ospelkaus.2006,Ulmanis.2016,Meister.2019}. This is a promising perspective to create quantum bubbles made possible by the method presented in this Letter.

An ideal shell-shaped BEC is point symmetric with respect to its center. Consequently, we asses the influence of shell-opening effects by comparing the density maxima $n_\text{max}(\theta,\varphi)$ and $n_\text{max}(\pi-\theta,\varphi+\pi)$ of the $^{41}$K ground state $|\psi_K(\mathbf{x})|^2$ along two opposing directions, characterized by the spherical angles $\theta$ and $\varphi$. Maximizing the difference along all possible directions gives the asymmetry
\begin{equation}
A = \max_{\{\theta,\varphi\} } \left| \frac{n_\text{max}(\theta,\varphi) - n_\text{max}(\pi - \theta,\varphi + \pi)}{n_\text{max}(\theta,\varphi) + n_\text{max}(\pi - \theta,\varphi + \pi)} \right|,
\label{eq:asymmetry}    
\end{equation}
which is a measure of how far the shell deviates from the ideal case. A perfect shell yields $A = 0$, whereas the opposite case of a completely opened-up shell gives rise to $A = 1$ because there is at least one direction along which $n_\text{max}(\theta,\varphi) = 0$.

\begin{figure}[t]
	\centering
	\includegraphics[width=180.4067pt,height=276.3105pt]{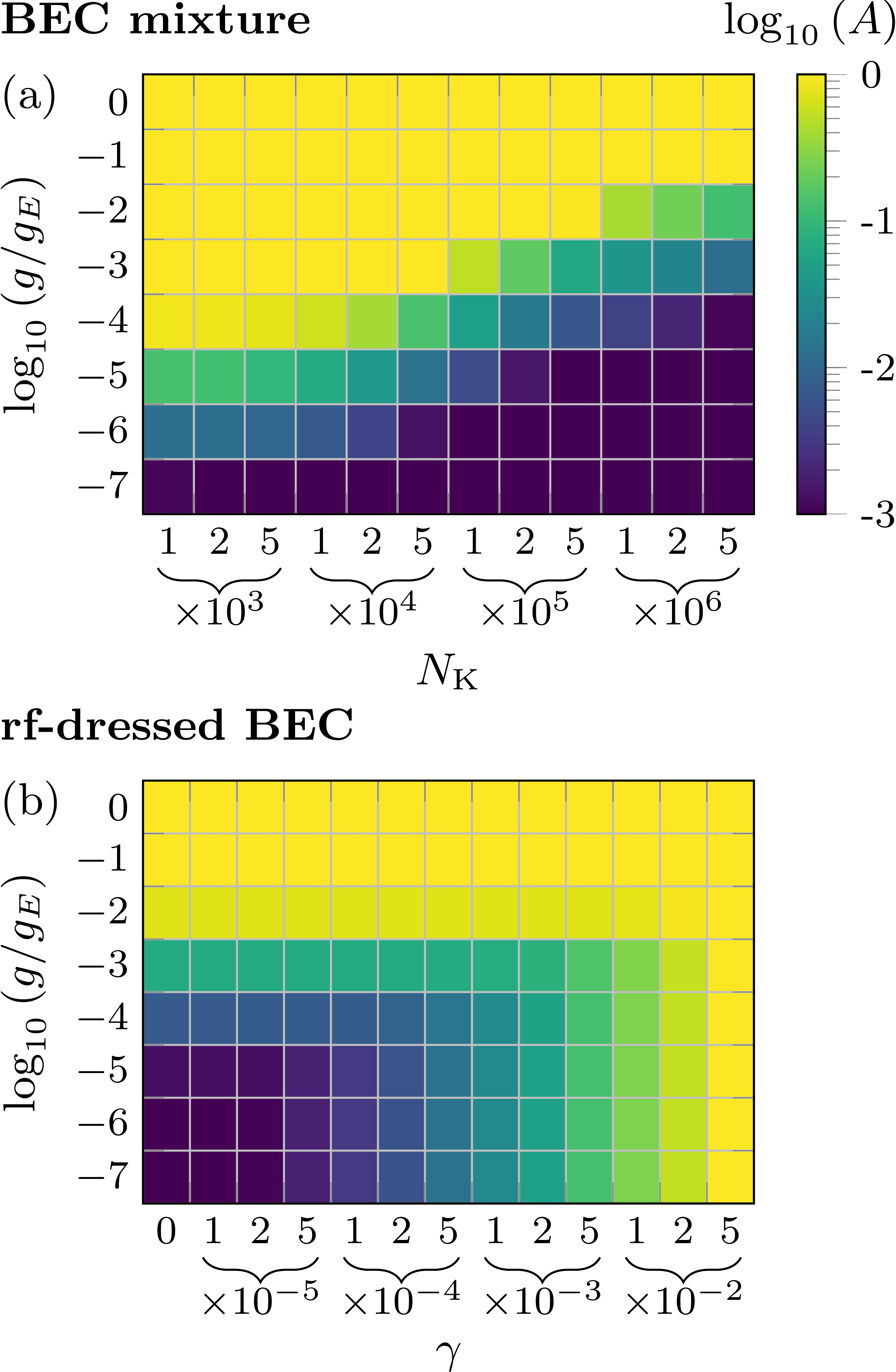}
	\caption{
	    Asymmetry $A$, Eq.~\eqref{eq:asymmetry}, for a BEC mixture (a) and an rf-dressed BEC (b) as a function of the gravitational acceleration $g$ ($g_\text{E} = 9.81~\text{m/s}^2$) and the number of potassium atoms $N_\text{K}$ or a linear gradient $\gamma$ of the Rabi frequency acting perpendicular to gravity, respectively. The cases $A=0$ and $A=1$ correspond to symmetrically filled and opened-up shells. (a) The mixture-based shell opens up for increasing $g$ due to differential gravitational sag, which can be partially compensated by larger atom numbers. (b) The rf-dressed shell opens up both for increasing gravity and Rabi frequency inhomogeneity. The parameters of the rf case~\cite{Note2} are chosen such that for $\gamma = g = 0$ the ground state density resembles the one for the mixture reference case shown in Fig.~\ref{fig:groundStates}.
	    }
	\label{fig:feasability}
\end{figure}

Figure~\ref{fig:feasability}(a) shows that gravity indeed prevents the creation of shell-shaped BECs in Earth-based laboratories for our reference case (yellow).
Only for considerably smaller gravitational accelerations, $g \leq 10^{-5} g_\text{E}$, does the system form an almost ideal shell (dark blue). An important aspect influencing this threshold is the available number of particles. More precisely, higher numbers of $^{41}$K atoms lead to an increased robustness against shell-opening, as exemplified in Fig.~\ref{fig:feasability}(a), because the width and density of the shell become larger.

In rf-dressed systems gravity essentially prevents the creation of shells in Earth-based laboratories due to the atoms pooling at the bottom of the shell-creating potential, opening the BEC up at the top~\cite{Colombe.2004}. In addition to gravity, rf-dressing is very sensitive to an inhomogeneity of the rf-field and non-perfect alignment with the static magnetic field. Both issues result in a spatially dependent Rabi frequency, which affects the depth of the shell potential~\cite{Lundblad.2019} forcing the atoms toward positions of lower Rabi frequency.

To compare shell-opening effects in mixture and rf-dressed shells, we consider $10^5$ atoms of $^{41}$K in a conventional \emph{bubble trap}~\cite{Zobay.2001} with parameters reproducing similar shell radius and thickness as our reference case, shown in Fig.~\ref{fig:groundStates}~\cite{Note2}. In addition to the gravitational potential $V_{g,\text{K}}(z)=m_K g z$, we model the position-dependent Rabi frequency in the bubble trap by a linear function $\Omega(x) = \Omega_0 \left(1 + \gamma x /x_0\right)$. Here $x_0$ is the position of the potential minimum along the $x$-axis and at $\mathbf{x}=(\pm x_0,0,0)$ the Rabi frequency has a relative deviation of $\gamma$ compared to its value $\Omega_0$ at $x=0$. A $\gamma > 0$ tilts the potential towards negative $x$-direction.

Figure~\ref{fig:feasability}(b) displays how an rf-dressed shell gradually opens up due to either gravity or a gradient in the Rabi frequency. Although both effects are completely decoupled in our simplified model, which generally is not the case, we can give estimates to create reasonably filled shells of the same size as our mixture-based reference case. Obviously, microgravity conditions are required as well as rather homogeneous Rabi frequencies with $\gamma \leq 10^{-3}$. Current experiments operate above this threshold with $\gamma_\text{exp} \approx 5 \times 10^{-3}$ for a trap diameter of $71~\text{\textmu m}$~\cite{Lundblad.2019, Carollo.2021}, where the degree of symmetry in the ground states strongly depends on the number of particles in the BEC.

\noindent\textit{Summary}.-- We have proposed an alternative method to create spherically symmetric shell-shaped BECs with dual-species atomic mixtures in a microgravity environment. Similar to the conventional rf-dressing scheme, both the ground state and the collective excitations identify the topological transition from a filled sphere to a shell-shaped BEC. Moreover, the shell structure of our mixture is conserved by the repulsive inter-species interaction during free expansion, allowing for a magnification of the dynamics on the shell. Additionally, we have quantified the effects of gravity and the atom number on the feasibility of achieving symmetrically filled shells and contrasted our results to the case of rf-dressing. 

We emphasize that our scheme based on dual-species mixtures has straightforward applications to related research areas, e.g., few-body physics in mixed dimensions~\cite{Nishida.2008,Lamporesi.2010}, as well as spinor~\cite{Kawaguchi.2012,StamperKurn.2013} and dipolar~\cite{Adhikari.2012,Diniz.2020,Arazo.2021} BECs on curved manifolds. 

\begin{acknowledgments}
This project is supported by the German Space Agency (DLR) with funds provided by the Federal Ministry for Economic Affairs and Energy (BMWi) due to an enactment of the German Bundestag under Grant Nos. 50WP1705 (BECCAL), 50WM1862 (CAL), and 50WM2060 (CARIOQA).
A.B. acknowledges funding provided by the Institute of Physics Belgrade, through the grant by the Ministry of Education, Science, and Technological Development of the Republic of Serbia.
N.G. acknowledges the support of the Deutsche Forschungsgemeinschaft (DFG, German Research Foundation) within the project A05 of CRC 1227 (DQmat) and under Germany’s Excellence Strategy – EXC-2123 QuantumFrontiers – 390837967. 
The authors are thankful for support by the state of Baden-W\"urttemberg through bwHPC and the German Research Foundation (DFG) through Grant no. INST 40/575-1 FUGG (JUSTUS 2 cluster).
\end{acknowledgments}

A.W. and P.B. contributed equally to this work.

\bibliography{bibliography}

\clearpage

\setcounter{equation}{0}
\renewcommand{\theequation}{S.\arabic{equation}}

\setcounter{figure}{0}
\renewcommand{\thefigure}{S\arabic{figure}}

\setcounter{table}{0}
\renewcommand{\thetable}{S\arabic{table}}

\onecolumngrid

\begin{center}
	{\normalfont\large\bfseries\centering {\it Supplemental material for}\\ 
	Shell-shaped Bose-Einstein condensates realized with dual-species mixtures}\\[23pt]	
\end{center}

\twocolumngrid

\section{Bogoliubov-de Gennes equations}
In this Letter we use the Bogoliubov-de Gennes equations (BdGEs) to describe collective excitations and to obtain the corresponding spectrum. Both the GPE~\eqref{eq:GPE} and the BdGEs
\begin{align}
    E u_{\alpha} & =  L_\alpha u_{\alpha} + \sum_\beta g_{\alpha\beta} \left[|\psi_\beta|^2 u_{\alpha} + |\psi_\alpha||\psi_\beta| (u_{\beta} + v_{\beta})\right]\nonumber\\
    -E v_{\alpha} & =  L_\alpha^\ast v_{\alpha} + \sum_\beta g_{\alpha\beta} \left[|\psi_\beta|^2 v_{\alpha} + |\psi_\alpha||\psi_\beta| (u_{\beta} + v_{\beta})\right]\label{eq:BdGEs}
\end{align}
for component $\alpha$ of a multi-component BEC can be derived by extending the calculation presented in Ref.~\cite{Ueda.2010}. Here we have suppressed the spatial arguments in the ground state solution $\psi_\alpha(\mathbf{x})=|\psi_\alpha(\mathbf{x})|e^{iS_\alpha(\mathbf{x})}$ of Eq.~\eqref{eq:GPE}, the quasiparticle mode functions $u_\alpha(\mathbf{x})$, $v_\alpha(\mathbf{x})$ as well as the linear operator 
\begin{equation}
    L_\alpha(\mathbf{x}) = -\frac{\hbar^2}{2 m_\alpha} [\nabla +i \nabla S_\alpha(\mathbf{x})]^2 + V_\alpha(\mathbf{x}) - \mu_\alpha.
\end{equation}
The chemical potential $\mu_\alpha$ is obtained together with the ground state by solving the time-independent GPE numerically. Since the wave functions of the ground states considered throughout this Letter are real, $S_\alpha$ are constants, $\nabla S_\alpha(\mathbf{x}) = 0$, and $L_\alpha^\ast = L_\alpha$.

The BdGEs~\eqref{eq:BdGEs} are an eigenvalue problem for the quasiparticle mode functions $\{u_\alpha,v_\alpha\}$ and the corresponding energies $E = \hbar \omega$. The low-frequency modes of the BdGEs describe collective excitations~\cite{Dalfovo.1999}, with $\omega$ being the frequency of the corresponding density oscillations. Everywhere in the supplemental material we suppress mode indices. For more details on the BdGEs we refer to Refs.~\cite{Dalfovo.1999,Ueda.2010}.

For our discussions of collective excitations, all potentials and ground states are spherically symmetric. This enables us to perform a separation of variables in the BdGEs by expanding the angular parts in terms of spherical harmonics. The BdGEs thus reduce to a system of linear one-dimensional differential equations with respect to the radial coordinate $r$, including a centrifugal potential $\hbar^2 l(l+1)/(2m_\alpha r^2)$. In Figs.~\ref{fig:breathing} and~\ref{fig:SupMat_breathing} (solid lines) we show the first few (positive) mode frequencies $\omega$ corresponding to $l=0$, with the parameters given in Tab.~\ref{tab:SupMat_reference} and excluding Goldstone modes.

\section{Numerical simulations}
In this Letter we have performed three different types of numerical simulations: (i) finding the ground state solution of the GPE, (ii) propagating the ground state wave function in time, and (iii) solving the BdGEs. To find the ground states in either one or three dimensions on a discretized grid, we have used the imaginary-time and split-step methods, giving rise to the results presented in Figs.~\ref{fig:groundStates},~\ref{fig:feasability} and~\ref{fig:SupMat_groundStates}~\cite{Vudragovic.2012,Sataric.2016}. A similar simulation but with real time can be used to propagate the ground state wave function in time, which enables us to consider the free expansion scenarios shown in Figs.~\ref{fig:expansion} and~\ref{fig:SupMat_expansion}. Additionally, the collective excitation frequencies can be accessed by direct simulation of the GPE~\eqref{eq:GPE} and performing a Fourier transformation of a quantity such as the expectation value and variance of the radial coordinate over time. These results are presented by dots in Figs.~\ref{fig:breathing} and~\ref{fig:SupMat_breathing}. Finally, for solving the BdGEs we use an eigensolver based on the finite element method.

\section{Comparison with rf-dressing approach}
To compare the mixture-based scheme with the rf-dressed one, we employ the rf-dressed potential~\cite{Zobay.2001,Lesanovsky.2006}
\begin{equation}
	V_\text{rf}(\mathbf{x}) = \frac{M_F g_F}{|g_F|} \sqrt{\left(\frac{m \omega_{0,\text{rf}}^2}{2 F} \mathbf{x}^2 - \hbar \Delta\right)^2 + (\hbar \Omega_0)^2},
	\label{eq:rfPotential}
\end{equation}
where $M_F$ denotes the projection of the total momentum $F$ of a dressed state in the hyperfine manifold with corresponding Land\'e factor $g_F$. The trap frequency $\omega_{0,\text{rf}}$ of the static magnetic trap is chosen such that the potential of the highest trapped bare state is given by $V_\text{st}(\mathbf{x}) = m \omega_{0,\text{rf}}^2 \mathbf{x}^2/2$. Moreover, $\Delta$ is the detuning of the rf-field with respect to the transition between neighboring bare states at the center of the trap, and $\Omega_0$ is the corresponding Rabi frequency. Here we only consider single-component BECs in rf-dressed traps and therefore drop all component-related indices.

\subsection{Ground states}
In order have a fair comparison between the shells created in both systems, we simulate the rf-dressed scheme with the same amount of $^{41}$K atoms as in the mixture case. A shell-shaped BEC in the mixture-based approach results from the combination of the harmonic trapping potential and the repulsion provided by the inner $^{87}$Rb core, giving rise to an effective potential $V_\text{eff}(\mathbf{x}) = g_\text{Rb,K}|\psi_\text{Rb}(\mathbf{x})|^2 + m_\text{K} \omega_{0,\text{K}}^2 \mathbf{x}^2/2$ for $^{41}$K. To have an rf-dressed shell with the same geometrical parameters, we fit the potential $V_\text{rf}$, Eq.~\eqref{eq:rfPotential}, to $V_\text{eff}(\mathbf{x})$ and thereby obtain the corresponding values for $\Delta$ and $\omega_{0,\text{rf}}$. All other parameters are chosen beforehand and the complete set of parameters is listed in Tab.~\ref{tab:SupMat_reference}.
\begin{table}[htb]
	\caption{\label{tab:SupMat_reference}%
	Parameters for the comparison of the two schemes resulting in shells of the same species and with the same geometrical sizes. Note that the parameters for the BEC mixture are the same as in Tab.~\ref{tab:reference}.
	}
	\begin{ruledtabular}
		\begin{tabular}{clcD{.}{.}{3.1}ccc}
			& \multicolumn{1}{c}{Species} & $N$ & \multicolumn{1}{c}{\begin{tabular}[c]{@{}c@{}}$\omega_0/(2\pi)$\\ (Hz)\end{tabular}}	& \begin{tabular}[c]{@{}c@{}} $a_{\text{Rb,K}}$\\ ($a_0$)\end{tabular}  & \begin{tabular}[c]{@{}c@{}} $\Delta/(2\pi)$\\ (kHz)\end{tabular} & \begin{tabular}[c]{@{}c@{}} $\Omega_0/(2\pi)$\\ (kHz)\end{tabular} \\\colrule
			\rule{0pt}{3.7ex} \begin{tabular}[c]{@{}c@{}}rf-dressed\\BEC\end{tabular}  & $^{41}$K\footnotemark[1]$^{,}$\footnotemark[2] & $10^5$ & 152.3 & & 3.78 & 2.5\\\colrule
			\rule{0pt}{2.5ex} \multirow{2}{*}{\begin{tabular}[c]{@{}c@{}}BEC\\mixture\end{tabular}} & $^{41}$K\footnotemark[2] & $10^5$ & 70 &  \multirow{2}{*}{85} & &\\
			& $^{87}$Rb\footnotemark[3] & $10^6$ & 51.3 & & &
		\end{tabular}
	\end{ruledtabular}
	\footnotetext[1]{The atoms are prepared in the $\lvert F = 2, M_F = 2  \rangle$ dressed state with $g_F=|g_F|$.}
	\footnotetext[2]{$a_\text{K,K} = 60~a_0$}
	\footnotetext[3]{$a_\text{Rb,Rb} = 100~a_0$}
\end{table}

\begin{figure}[htb]
	\centering
	\includegraphics[width=245.99841pt,height=228.28574pt]{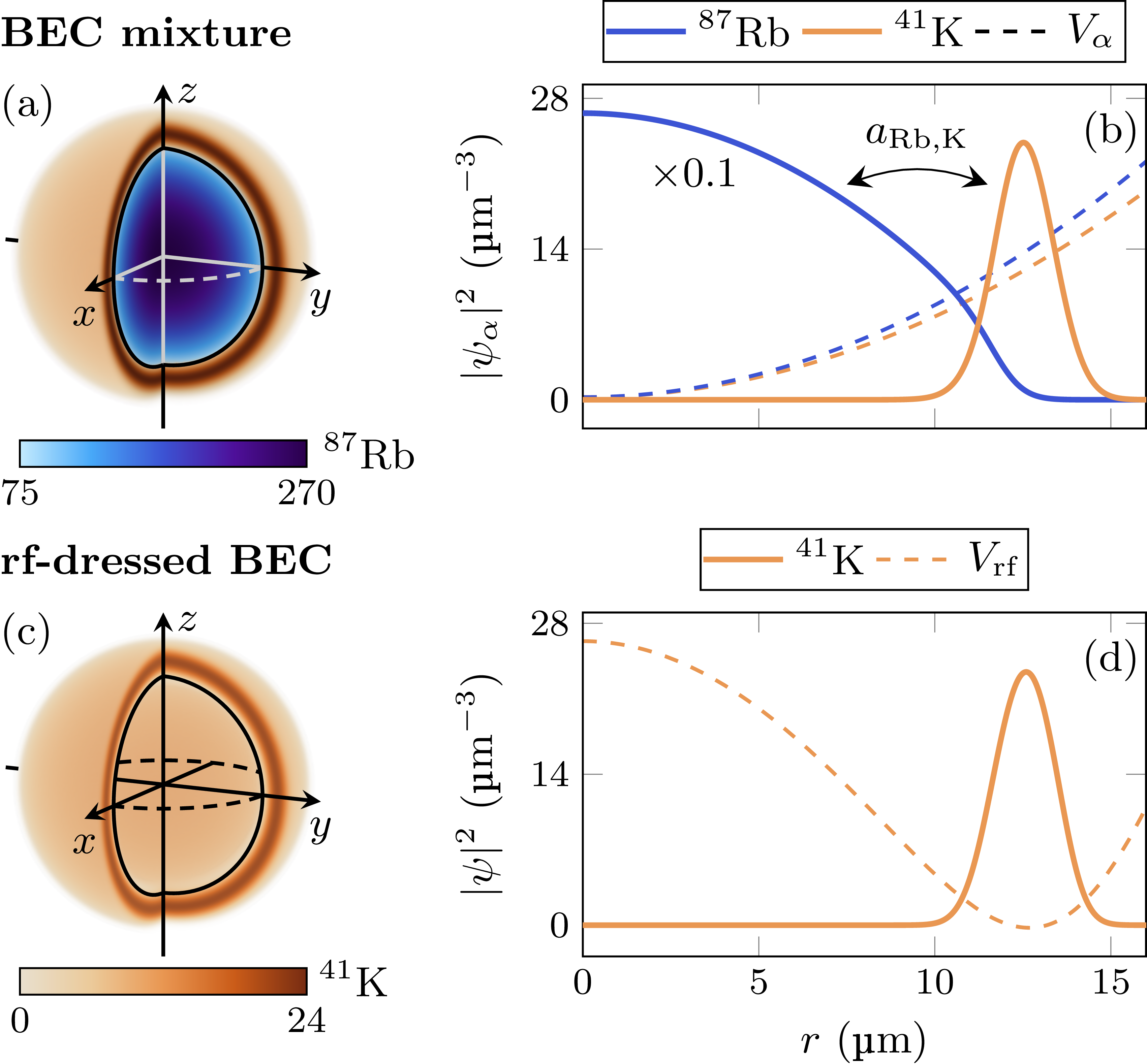}
	\caption{\label{fig:SupMat_groundStates}%
	Spherically-symmetric shell-shaped ground-state density distributions $|\psi_\alpha|^2$ of a $^{87}$Rb-$^{41}$K BEC mixture (a),(b) and an rf-dressed $^{41}$K BEC (c),(d) using the parameters presented in Tab.~\ref{tab:SupMat_reference}. (a),(c) Cut-open three-dimensional density plots (colorbar units in $\text{\textmu m}^{-3}$). (b),(d) Density profiles (solid lines) along the radial direction and corresponding trapping potentials (dashed lines) $V_\alpha(\mathbf{x}) = m_\alpha \omega_{0,\alpha}^2 \mathbf{x}^2/2$ and $V_\text{rf}(\mathbf{x})$, Eq.~\eqref{eq:rfPotential}. The mixture forms a shell due to the interplay between the harmonic confinements and a repulsive inter-species interaction governed by the $s$-wave scattering length $a_\text{Rb,K}$. In contrast, the rf-dressed shell-shaped BEC relies on the trapping potential being a double-well in all directions.
	}
\end{figure}

Figure~\ref{fig:SupMat_groundStates} presents the ground states of both systems for the parameters given in Tab.~\ref{tab:SupMat_reference} and clearly shows that the ground-state density distributions of $^{41}$K (orange) in the BEC mixture, Fig.~\ref{fig:SupMat_groundStates}(b), and the rf-dressed BEC, Fig.~\ref{fig:SupMat_groundStates}(d), are almost identical.

\subsection{Collective excitation spectrum}
\begin{figure}[htb]
	\centering
    \includegraphics[width=245.99826pt,height=278.21284pt]{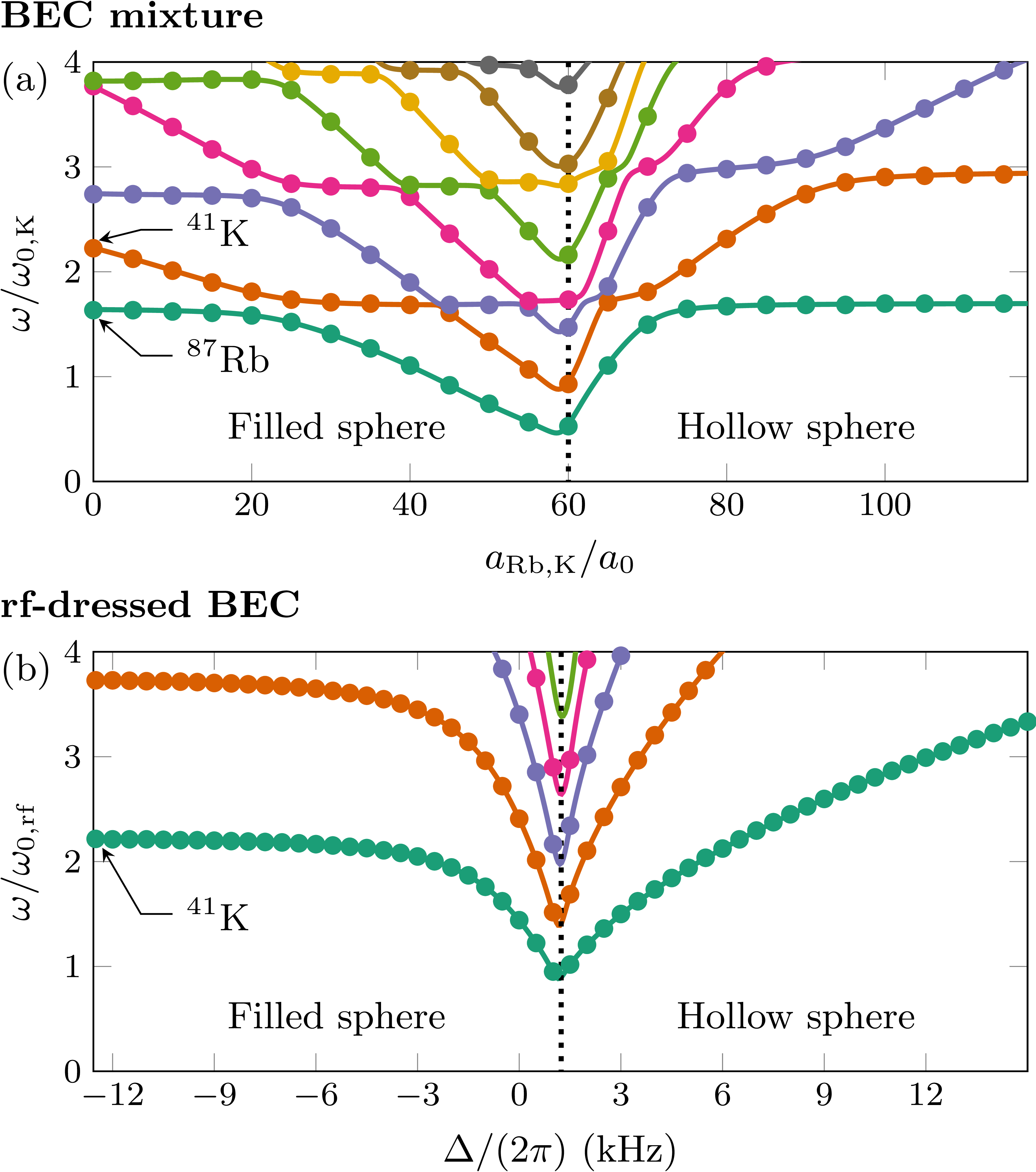}
	\caption{\label{fig:SupMat_breathing}%
	Mode frequency $\omega$ of the lowest-lying $l=0$ collective excitations for a $^{87}$Rb-$^{41}$K BEC mixture (a) and an rf-dressed $^{41}$K BEC (b) as a function of the inter-species scattering length $a_\text{Rb,K}$ and detuning $\Delta$, respectively. The solid lines and dots are determined by the solutions of the BdGEs~\eqref{eq:BdGEs} and numerical simulations of the GPEs~\eqref{eq:GPE}, accordingly. In both systems the common minimum of the frequencies is a clear sign of the hollowing transition marked by the dotted vertical lines, where $|\psi_\text{K}(0)|^2/\mathrm{max}|\psi_\text{K}(\mathbf{x})|^2$ drops below $10^{-2}$. Moreover, in the respective harmonic limit, where both systems reduce to single-component BECs in spherically symmetric harmonic traps, the excitation frequencies are given by Eq.~\eqref{eq:tfSpectrum} and the lowest frequencies are marked accordingly.
    }
\end{figure}

In Fig.~\ref{fig:SupMat_breathing} we show the collective excitation frequencies for the first few $l=0$ modes as a function of $a_\text{Rb,K}$, (a), or $\Delta$, (b). In the rf-dressed case, this spectrum has a minimum at a certain $\Delta$ corresponding to the transition between a filled sphere and a hollow sphere~\cite{Padavic.2017,Sun.2018}. In the mixture case, this spectrum displays a similar feature albeit at a certain value of $a_\text{Rb,K}$.

Further similarities can be seen in the limit of the harmonically trapped single-component BEC. In the mixture scheme, this limit corresponds to $a_\text{Rb,K} = 0$, where both components are completely decoupled, and are reduced to the well-studied case of a BEC in a spherically symmetric harmonic trap. The collective excitation spectrum
\begin{equation}
    \omega_\alpha = \omega_{0,\alpha} \sqrt{2n^2+2nl+3n+l}
    \label{eq:tfSpectrum}
\end{equation}
can be obtained analytically using the Thomas-Fermi approximation~\cite{Ueda.2010}.

Hence, the spectrum depends on the two numbers $n=0,1,2,\ldots$ and $l=0,1,2,\ldots$ as well as the corresponding harmonic trap frequency $\omega_{0,\alpha}$. In Fig.~\ref{fig:SupMat_breathing}(a) we mark the first nonzero excitation frequency for $^{41}$K and $^{87}$Rb, respectively. At $a_\text{Rb,K} = 0$ all shown excitation frequencies agree with Eq.~\eqref{eq:tfSpectrum} and can be matched to either of the components.

In the rf-dressed scheme, the harmonic limit corresponds to $\Delta < 0$ with $\Omega_0/|\Delta| \ll 1$, where the potential, Eq.~\eqref{eq:rfPotential}, reduces to a harmonic potential with an offset
\begin{equation}
    V_\text{rf}(\mathbf{x}) \approx - \frac{M_F g_F}{|g_F|} \hbar \Delta +  \frac{M_F g_F}{|g_F|} \frac{m \omega_{0,\text{rf}}^2}{2F} \mathbf{x}^2 + \mathcal{O}\left(\frac{\Omega_0^2}{\Delta}\right).
\end{equation}
In this limit, the collective excitations of the highest trapped state with $M_F g_F = |g_F|F$ are thus also described by Eq.~\eqref{eq:tfSpectrum}. Due to the scaling in the respective trap frequency in Fig.~\ref{fig:SupMat_breathing}, the excitation frequencies of the rf-dressed BEC, Fig.~\ref{fig:SupMat_breathing}(b), start at the same value as the excitation frequencies of $^{41}$K in the mixture, Fig.~\ref{fig:SupMat_breathing}(b).

Let us now consider the behavior as we move away from the respective harmonic limit by looking at the lowest excitation frequencies, as marked in Fig.~\ref{fig:SupMat_breathing}. In the mixture scheme, increasing the interaction between the components leads to an immediate decrease of the frequencies belonging to $^{41}$K. In contrast, the excitation frequencies of $^{87}$Rb stay almost constant. This can be explained by the fact that the ground state of $^{87}$Rb barely changes while $^{41}$K ultimately transforms into a shell which in turn is due to the large disparity between the particle numbers of both components. A consequence of these two different behaviors of the excitation frequencies is the display of avoided crossings. In the rf-dressed scheme, a second component is absent. Furthermore, the behavior of the excitation frequencies is different to those of $^{41}$K in the mixture scheme, as they start to decrease only slowly. This is due to the fact that the potential is deformed away from its harmonic form rather slowly and it is only obtaining a double-well-like structure for $\Delta > 0$.

\subsection{Free expansion}
\begin{figure}[htb]
	\centering
	\includegraphics[width=246.0pt,height=210.63333pt]{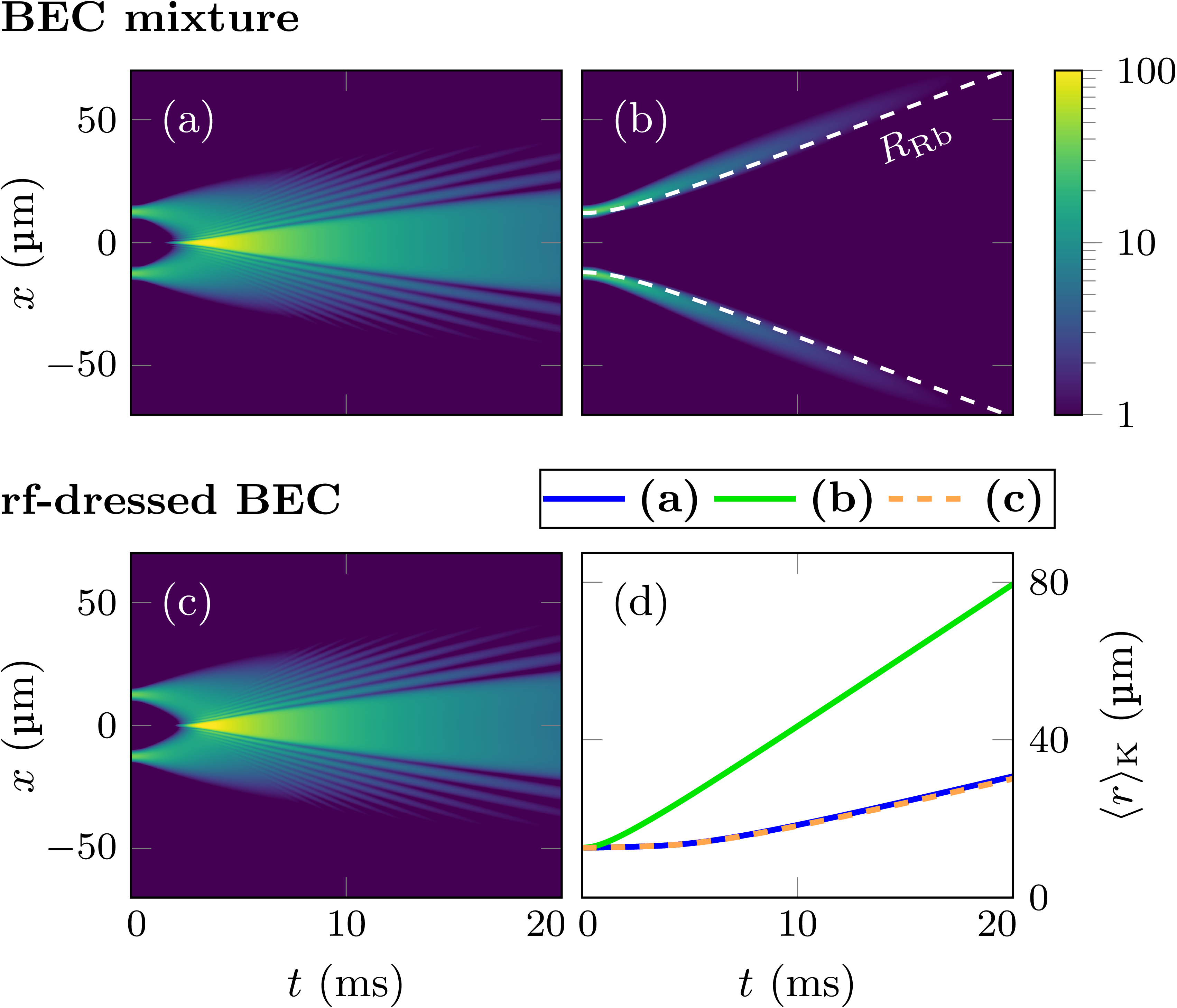}
	\caption{\label{fig:SupMat_expansion}%
	Free expansion of the spherically symmetric density distributions $|\psi_\text{K}|^2$, which are initially prepared in the form presented in Fig.~\ref{fig:SupMat_groundStates}, along the $x$-direction for a $^{87}$Rb-$^{41}$K BEC mixture (a),(b) and an rf-dressed $^{41}$K BEC (c) (colorbar units in $\text{\textmu m}^{-3}$). (a) By switching off both the confinement and the interaction between the two species, the shell can expand inwards until it reaches the center and shows a self-interference pattern. (b) Leaving the interaction at $a_\text{Rb,K} = 85~a_0$ leads to an expanding shell with its size being proportional to the edge of the expanding inner rubidium core $R_\text{Rb}$ defined by $|\psi_\text{Rb}|^2$ dropping below $10^{-2}$ of its peak value. (c) Switching off all magnetic fields in the rf-dressed BEC results in a similar free expansion as the mixture case in (a). (d) Tracking the expectation value $\langle r\rangle_\text{K}$ over time reveals the similarity between (a) and (c) as well as the increasing radius of the expanding shell in (b).
	}
\end{figure}

In Fig.~\ref{fig:SupMat_expansion} we compare the free expansions of both ground states displayed in Fig.~\ref{fig:SupMat_groundStates}. This is done by switching off the optical dipole trap in the mixture scheme and both the static magnetic and rf field in the rf-dressed scheme. Unsurprisingly, if the interaction between $^{87}$Rb and $^{41}$K is additionally tuned to zero, $a_\text{Rb,K} = 0$, Fig.~\ref{fig:SupMat_expansion}(a), the free expansion of the density distribution $|\psi_\text{K}|^2$ is similar to the one reported for the rf-dressed BEC~\cite{Lannert.2007,Tononi.2020}, Fig.~\ref{fig:SupMat_expansion}(c), due to the fact that both ground states have almost identical shells. In contrast, at $a_\text{Rb,K} = 85~a_0$, Fig.~\ref{fig:SupMat_expansion}(b), the free expansion of $|\psi_\text{K}|^2$ is completely different and features an expanding shell.
Moreover, for the expansion scenarios shown in Fig.~\ref{fig:SupMat_expansion}(a)-(c), we track the time-dependent expectation value $\langle r\rangle_\text{K} = 4\pi\int_0^\infty \mathrm{d}r~r^3 |\psi_\text{K}(\mathbf{x},t)|^2$ of the radial coordinate for $^{41}$K and present the results in Fig.~\ref{fig:SupMat_expansion}(d). Here we see the similarity of the two expansion dynamics displayed in Fig.~\ref{fig:SupMat_expansion}(a) and (c) as well as the much faster expanding shell scenario in Fig.~\ref{fig:SupMat_expansion}(b).

\subsection{Feasibility}
Finally, in order to compare shell-opening effects in both schemes, Fig.~\ref{fig:feasability}, we modify the Rabi frequency in Eq.~\eqref{eq:rfPotential} and replace $\Omega_0$ by $\Omega(x) = \Omega_0 \left(1 + \gamma x /x_0\right)$. In this way, we can model the inhomogeneity of the Rabi frequency. Here $x_0 = \sqrt{2F\hbar\Delta/(m \omega_{0,\text{rf}}^2)}$ is the position of the minimum of the potential $V_\text{rf}(\mathbf{x})$, Eq. \eqref{eq:rfPotential}, along the $x$-direction. Using the parameters listed in Tab.~\ref{tab:SupMat_reference} and the displayed values for $g$ and $\gamma$, we calculate first the ground state of the system and afterwards the asymmetry $A$, Eq.~\eqref{eq:asymmetry}.

\clearpage

\end{document}